\documentclass[aps, pre, reprint, groupedaddress, superscriptaddress, amsmath, amssymb, showpacs]{revtex4-2}
\usepackage{bm}
\usepackage{graphicx}
\graphicspath{{fig/}}

\begin{document}
    \title{Simulation of ultracold plasma expansion in homogeneous magnetic field}

    \author{E. V. Vikhrov}
    \email{vikhrov-e@ihed.ras.ru}
    \author{B. B. Zelener}
    \author{B. V. Zelener}
    \affiliation{Joint Institute for High Temperatures of the Russian Academy of Sciences, Izhorskaya St. 13, Bldg. 2, Moscow 125412, Russia}

    \begin{abstract}
        We present molecular dynamics simulation results for an ultracold $^{40}$Ca plasma in a constant, homogeneous magnetic field. 
        The magnetic field induces a significant spatial separation of charges. 
        Furthermore, a significant fraction of the electrons escapes from the region bounded by the ions, in contrast to the case of plasma expansion into a vacuum. 
        This leads to the formation of a quasineutral plasma core containing the remaining electrons and surrounded by an outer, thick ion shell, where the ion density drops sharply. 
        Anisotropy in the ion kinetic energy is also observed, depending on the direction of ion motion relative to the magnetic field lines. 
        The simulation results are in agreement with experimental data on ion dynamics.
    \end{abstract} 

    \maketitle
    \section{Introduction}
            The behavior of plasma in magnetic fields has long attracted considerable scientific interest. 
        There have been several experiments on the expansion of ion clouds in Earth's magnetic field, carried out in the ionosphere and at greater heights \cite{Pilipp1971}.

            Experiments on the expansion of ultracold plasmas in a constant, homogeneous magnetic field have also been carried out under laboratory conditions 
        \cite{bib_zhang2008,bib_sprenkle2022,bib_pak2024preliminary}. These experiments provide insight into ion dynamics and allow various theoretical approaches to describing this behavior to be tested. 
        However, several important questions remain unresolved. In particular, there is a lack of experimental data on electron dynamics due to experimental limitations.

            In this paper, we present the results of molecular dynamics simulations of an ultracold Ca plasma in a constant, homogeneous magnetic field. 
        The simulations show that the magnetic field induces significant charge separation even at low magnetic field strengths 
        Moreover, the charge imbalance may reach significant values compared with those observed during ultracold plasma expansion into a vacuum.  
        During plasma expansion, an outer, 
        thick ion shell with low density is formed. This shell surrounds and confines a quasineutral plasma core containing the remaining electrons. 
        Another notable effect is the distortion of the initial ion velocity distribution, while the electron distribution remains Maxwellian. 
        Anisotropy in the ion kinetic energy is also observed, depending on the direction of ion motion relative to the magnetic field lines. 
        Accurate calculations of ion chaotic motion and ion fluorescence spectra are carried out.
          
    \section{Simulation Details}\label{sect_sim_det}
            The simulation parameters are chosen with a focus on experimental conditions \cite{bib_zhang2008, bib_sprenkle2022, bib_pak2024preliminary}.

            We simulate $^{40}$Ca ultracold plasmas by means of molecular dynamics over time scales up to $10$~$\mu$s. The masses and charges of the particles are set equal to their real values for $^{40}$Ca$^+$ ions 
        and electrons respectively. The numbers of both types of particles are equal, $N = N_{\text{e}} = N_{\text{i}}$ and are varied within the range $N = 2.5\cdot10^4 \text{ to } 10^5$. The initial ion 
        temperature is set to 
        $T_{\text{i0}} = 10^{-3}$~K, while the initial electron temperature is varied within the range $T_{\text{e0}} = 10 \text{ to } 400$~K. The initial peak number density is set to $n_0 = 10^{9}$~cm$^{-3}$.
        The value of the initial plasma size $\sigma_0$ depends on $N$ and $n_0$ as $\sigma_0 = (N / n_0)^{1/3} / \sqrt{2\pi}$. An external constant homogeneous magnetic field is applied to the plasma
        along Cartesiam z-axis. The magnetic field strength is varied in range $B = 10 \text{ to } 10^3$~G. Other simulation details are presented in our previous papers 
        \cite{bib_vikhrov2020simulated, bib_vikhrov2021ion, bib_bronin2023ultracold, vikhrov2025diocotron}.

    \section{Comparison with the experimental data}
\begin{figure}[ht!]
            \includegraphics[width=\linewidth]{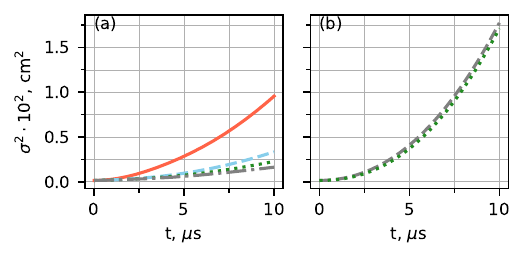}
            \caption{
                Dependency of $\sigma^2(t)$ for $n_0 = 10^9$~cm$^{-3}$, $T_\text{e0} = 100$~K, $N = 2.5\cdot10^4$:
                (a) perpendicular to the magnetic field lines for $B = 10$~G (solid), $B = 100$~G (dashed), $B = 250$~G (dotted) and $B = 500$~G (dash-dot);
                (b) along the magnetic field lines for $B = 0$~G (dashed) and $B = 100$~G (dotted).  
            }
            \label{fig_sigmas_com}
        \end{figure}

            In Fig. \ref{fig_sigmas_com}(a) the ion variance $\sigma^2(t)$ for various values of $B$ in the direction perpendicular to the magnetic field lines is shown.
            The ion variance grows more slowly as the magnetic field strength increases, which is in agreement with the experimental results \cite{bib_zhang2008, bib_sprenkle2022}. 
        From the data shown in Fig. \ref{fig_sigmas_com}(b), where the ion variance for $B=100$~G along the magnetic field lines is compared with that for expansion into a vacuum, it follows that the effect 
        of the magnetic field on the expansion in this direction is negligible. This result is also in agreement with the experimental results \cite{bib_zhang2008, bib_sprenkle2022}.

            The paper \cite{bib_sprenkle2022} highlights the difficulties encountered when obtaining the time dependence of $\sigma(t)$ for the case of expansion in a magnetic field using various 
        theoretical approaches. One can rewrite the dependence of $\sigma^2(t)$ given in \cite{killian2007ultracold} and modify it for the case of expansion perpendicular to the magnetic field lines as follows:
        \begin{equation}
            \label{eq_sigma}
            \sigma_\perp^2(t) = \sigma_0^2 + c_\text{s}^2t^2 - \frac{B^2}{m_\text{i}}\alpha t^2,
        \end{equation}
        where $\alpha$ is a fit parameter, and $c_s = \sqrt{k_\text{b}T_\text{e0} / m_\text{i}}$ is the ion-acoustic speed.

        The expression in Eq. (\ref{eq_sigma}) can be reduced to the form presented in \cite{bib_sprenkle2022}.

        Since the second term on the right-hand side of Eq. (\ref{eq_sigma}) is associated with the electron pressure, the third term is associated with the effective magnetic pressure, 
        which suppresses plasma expansion. The expression is in good agreement with the simulation data.
        \begin{figure}[ht!]
            \includegraphics[width=\linewidth]{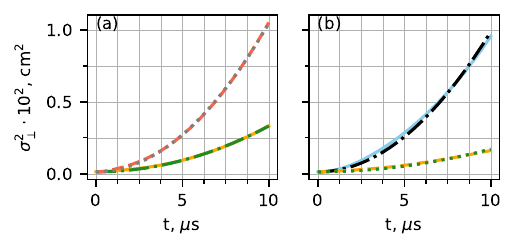}
            \caption{
                Dependency of $\sigma_\perp^2(t)$ for $n_0 = 10^9$~cm$^{-3}$, $N = 2.5\cdot10^4$:
                (a) at $B = 100$~G, simulation (solid) and fit (dashdot) for $T_\text{e0}=100$~K, simulation (dashed) and fit (dotted) for $T_\text{e0}=400$~K;
                (b) at $T\text{e0} = 100$~K, simulation (solid) and fit (dashdot) for $B=10$~G, simulation (dashed) and fit (dotted) for $B = 500$~;
            }
            \label{fig_sigmas_fit}
        \end{figure}
        This can be seen from the data in Fig. \ref{fig_sigmas_fit}, where $\sigma_\perp^2(t)$ is plotted for various values of $B$ and $T_\text{e0}$.

\begin{figure}[ht!]
            \includegraphics[width=0.52\linewidth]{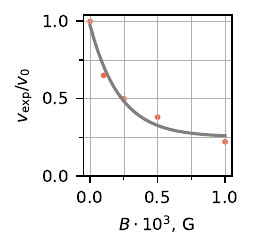}
            \caption{
                Dependency of $v_\text{exp} / v_0$ on $B$, fit(solid) vs simulation (circles).
            }
            \label{fig_vel_fit}
        \end{figure}

            The simulation results for the plasma expansion velocity, $v_\text{exp}(B) = \left. d\sigma_\perp/dt \right|_{t\rightarrow \infty}$, are also in good agreement with the exponential fit 
        proposed in \cite{bib_sprenkle2022}, as shown in Fig. \ref{fig_vel_fit}.

        Here the plasma expansion velocity into vacuum $v_0$ is used for normalization.

    \section{Charged shell formation}
\begin{figure*}[ht!]
            \includegraphics[width=\linewidth]{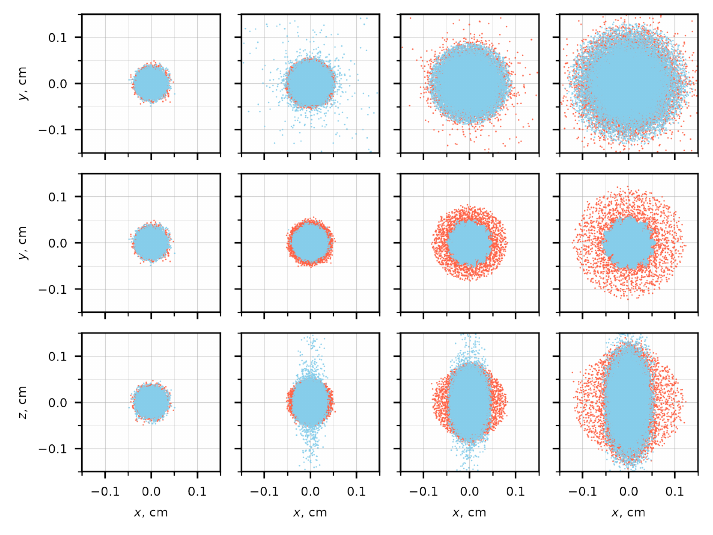}
            \caption{
                Projections of electrons (blue) and ions (red) at $B=0$~G (top, xy), compared to $B=100$~G (middle, xy) and $B=100$~G (bottom, xz) at various moments of time (from left 
                to right) $t = 0$~$\mu$s, $t = 3$~$\mu$s, $t = 6$~$\mu$s and $t = 9$~$\mu$s for $n_0=10^{9}$~cm$^{-3}$, $N = 2.5\cdot10^4$.
            }
            \label{fig_plasma_evol}
        \end{figure*}

            The simulations also provide insight into plasma particle dynamics. Within the simulated range of $B$ and initial electron temperatures $T_\text{e0}$, the plasma expansion exhibits a notable feature.
        The magnetic field induces significant charge separation and forms a thick outer ion layer in which the ion density drops sharply. 
        This layer surrounds a quasineutral plasma core containing the remaining electrons. 
        This can be seen in the second row of Fig. \ref{fig_plasma_evol}, where the projections of the plasma particles onto the $xy$ plane are plotted.

            Non-neutrality in the wings of the density distribution is known \cite{lyubonko2012collective} to occur in the case of plasma expansion into a vacuum. 
        This occurs only at the very edge of the plasma and is accompanied by the formation of an ion front \cite{bib_vikhrov2020simulated, bib_vikhrov2021ion, warrens2023shockwaves}. 
        In this case, the ions and electrons diffuse into the surrounding space at approximately the same rate, as can be seen in the upper row of Fig. \ref{fig_plasma_evol}.

        However, this is not the case when an external magnetic field is applied. Within the simulated range of $B$, the electrons are magnetized, whereas the ions are not. 
        This leads to different diffusion rates for the two charge species in the direction perpendicular to the magnetic field lines. 
        As a result, the plasma separates into an outer, thick shell of ions with low density and an inner quasineutral plasma core.

            The boundary of the shell diffuses outward across the magnetic field lines at a higher rate, while the quasineutral plasma core remains almost unchanged in size. 
        This can be seen in the middle row of Fig. \ref{fig_plasma_evol}. A clear difference between the expansion into a vacuum and the expansion perpendicular to the magnetic field lines 
        can be seen by comparing the upper and middle rows of Fig. \ref{fig_plasma_evol}.

        This finding complements the experimental data \cite{bib_zhang2008, bib_sprenkle2022}.

    \section{Spatial characteristics}
\begin{figure}[ht!]
            \includegraphics[width=\linewidth]{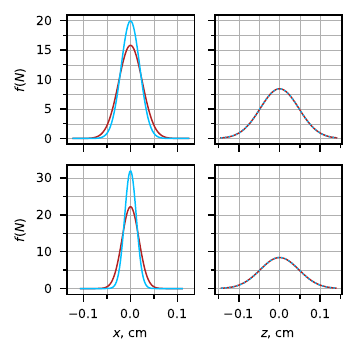}
            \caption{
                Spatial distributions of electrons (blue) and ions (red) for $N=2.5\cdot10^4$ and $n_0 = 10^9$~cm$^{-3}$ at $T_\text{e0} = 10$~K and $t = 10$~$\mu$s for various values of magnetic fields:
                $B = 100$~G (top), $B = 1000$~G (bottom).
                Left column corresponds to x distribution, right column to z distribution.
            }
            \label{fig_spat_distr_comp_te10}
        \end{figure}

            The simulations show that the spatial distribution of the ions remains unchanged, in agreement with the experimental results \cite{bib_zhang2008,bib_pak2024preliminary}. 
        Moreover, the same behavior is observed for the electrons within the region occupied by the ions. 
        This is demonstrated in Fig. \ref{fig_spat_distr_comp_te10}, where the spatial distributions of the plasma particles within the ion region at $t = 10$~$\mu$s are shown for various values of $B$.

            The left column of Fig. \ref{fig_spat_distr_comp_te10} shows the ion and electron distributions along the x direction. Although the distribution functions have the same shape, their central 
        moments differ. In particular, the ion variance is larger than the electron one.

            The distribution functions of the ions and electrons along the z direction are plotted in the right column of Fig. \ref{fig_spat_distr_comp_te10}. 
            Since the distributions almost completely overlap, the expansion along the z direction is self-similar, as also reported in \cite{bib_zhang2008,bib_pak2024preliminary}.

        \begin{figure}[ht!]
            \includegraphics[width=\linewidth]{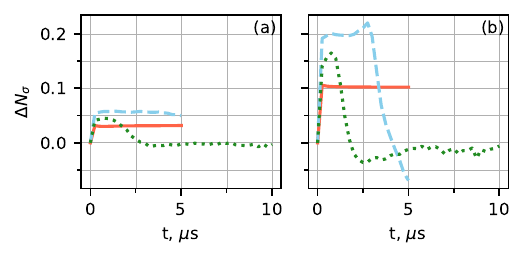}
            \caption{
                Dependencies of $\Delta N_\sigma (t)$ for $N=2.5\cdot10^4$ and $n_0~=~10^9$~cm$^{-3}$ in $r = 3\sigma_{\text{e}\perp}(t)$ at two different electron temperatures for various 
                magnetic field strength:
                (a)$B = 0$~G, (solid), $B = 10$~G, (dashed) and $B = 1000$~G, (dotted) at $T_\text{e0} = 10$~K;
                (b)$B = 0$~G, (solid), $B = 10$~G, (dashed) and $B = 1000$~G, (dotted) at $T_\text{e0} = 100$~K.
            }
            \label{fig_imb_sigm}
        \end{figure}

            The charge imbalance is $\Delta~N~=~(N_\text{i}^\prime~-~N_\text{e}^\prime)~/~N_\text{i}^\prime $ where $N_\text{i,e}^\prime$ is the number of corresponding particles 
        inside a given region. In the previous section, the quasineutrality of the inner plasma core was noted. 
        This follows from the calculated value of $\Delta N_\sigma$ inside a sphere of radius $3\sigma_{\text{e}\perp}(t)$, where $\sigma_{\text{e}\perp}(t)$ is the standard deviation 
        of the spatial electron distribution along the x direction.
        When defined in this way, the size of the region increases proportionally with the plasma expansion over time. Figure \ref{fig_imb_sigm} shows the dependence of $\Delta N_\sigma$ 
        for various values of $B$ and $T_\text{e0}$. It can be noted that the neutrality of the region, which is distorted by the initial electron evaporation, is restored more rapidly for larger values of $B$.
        It is also notable that this behavior is not observed for plasma expansion into a vacuum.

        \begin{figure}[ht!]
            \includegraphics[width=\linewidth]{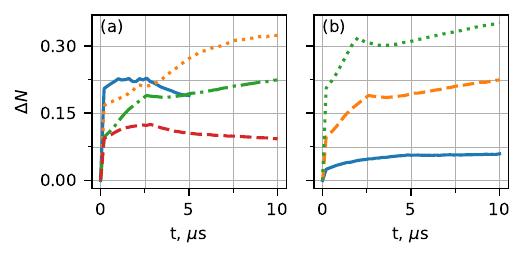}
            \caption{
                Dependencies of $\Delta N (t)$ for $N=2.5\cdot10^4$ and $n_0~=~10^9$~cm$^{-3}$:
                (a)$B~=~0$~G (solid), $B~=~10$~G (dotted), $B~=~100$~G (dashdot), $B~=~1000$~G (dashed) at $T_\text{e0} = 100$~K;
                (b)$T_\text{e0}~=~10$~K (solid), $T_\text{e0}~=~100$~K (dashed) and $T_\text{e0}~=~400$~K (dotted) at $B~=~100$~G.
            }
            \label{fig_imb_comparison}
        \end{figure}

            The magnetic field affects the way electrons leave the region bounded by the ions. The electrons leave this region only along the magnetic field lines. 
        The charge imbalance in this region is plotted in Fig. \ref{fig_imb_comparison}. It remains approximately constant only for $B = 0$~G and $B = 1000$~G. 
        For large values of $B$, $\Delta N(t)$ is suppressed compared with the case of expansion into a vacuum, as shown in Fig. \ref{fig_imb_comparison}(a). 
        For $B = 100$~G, $\Delta N(t)$ tends to increase. At low values of $B$, the largest fraction of electrons leaves the plasma. 
        This behavior differs from the case of ultracold plasma expansion in an external quadrupole magnetic field, where the charge imbalance is independent of the magnetic 
        field strength \cite{bib_bronin2023ultracold}. It is also notable that the magnetic field significantly suppresses the initial electron evaporation compared with the case of plasma expansion 
        into a vacuum.

            In Fig. \ref{fig_imb_comparison}(b), the dependence of $\Delta N(t)$ on time is shown for various values of $T_\text{e0}$ at fixed $B$. 
        The initial electron temperature remains a significant factor in determining the behavior of $\Delta N(t)$.

            Despite the initial suppression, the effect of the magnetic field on electron confinement within the plasma weakens over time, as also follows from the data shown in Fig. \ref{fig_imb_comparison}.

    \section{Plasma temperature and ion spectra}
        \begin{figure}[ht!]
            \includegraphics[width=\linewidth]{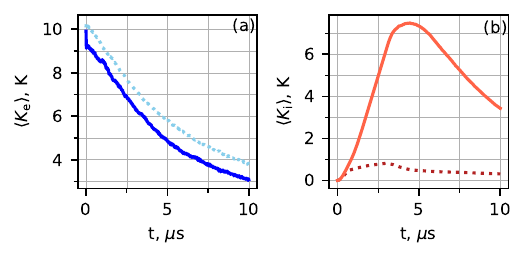}
            \caption{
                Time dependencies of kinetic energy $\langle K(t)\rangle$:
                (a) of electrons along(solid) and across (dotted) magnetic field lines; 
                (b) of ions along(solid) and across (dotted) magnetic field lines; 
                for $N=2.5\cdot10^4$, $n_0~=~10^9$~cm$^{-3}$, $B = 100$~G and $T_\text{e0} = 10$~K.
            }
            \label{fig_kin}
        \end{figure}

            Among other results, the experiment reported in \cite{bib_pak2024preliminary} provides data on the evolution of the ion temperature. 
        The authors of \cite{bib_pak2024preliminary} show that the ion temperature decreases as the initial electron temperature increases. 
        However, our simulations do not reproduce this result. Moreover, the simulations reveal several other features.

            Fig. \ref{fig_kin} shows the time dependence of the mean kinetic energies of ions and electrons, $\langle K_\text{i,e}\rangle$, calculated within a sphere of radius $3\sigma_0$.
    
            The initially uniform electron velocity distribution rapidly becomes Maxwellian and remains so over time. 
        Fig. \ref{fig_kin}(a) shows the time dependence of $\langle K_\text{e}(t)\rangle$ along the x and z directions. The effect of the magnetic field on $\langle K_\text{e}(t)\rangle$ is negligible.

            The initial Maxwellian ion velocity distribution becomes anisotropic over time. The behavior of $\langle K_\text{i}(t)\rangle$ strongly depends on the direction of ion motion. 
        Moreover, $\langle K_\text{i}(t)\rangle$ is larger along the z direction. 
        This is shown in Fig. \ref{fig_kin}(b) and is contrary to the experimental results reported in \cite{bib_pak2024preliminary}, where the ion temperature $T_\text{i}$ is isotropic.

            Based on the simulation data, the ion temperature $T_\text{i}$ can be estimated as the difference between the total kinetic energy and the kinetic energy associated with directed motion. 
        In order to calculate $T_\text{i}$, we construct a cubic mesh in the region occupied by the ions. 
        Since this region tends to expand over time, the cell size of the mesh is adjusted accordingly based on the value of $\sigma(t)$, while the number of cells remains unchanged.

            At each moment in time, the local ion temperature $T_q$ can be calculated for each cell $q$ as follows:
        \begin{equation}
            T_q(t) = \frac{m_\text{i}}{3k_\text{b}}\frac{1}{N_q}\sum\limits_{p \in q}|\bm{v}_p - \bm{u}_q|^2, 
            \label{eq_temp_mesh}
        \end{equation}
        where $p$ is ion number inside cell $q$, $N_q$ is total number of ions inside the cell, $\bm{v}_p$ is velocity of ion, and $\bm{u}_q$ the velocity of local flow inside the cell.

            The spatial distribution of the ion temperature can be reduced to a single scalar value:
        \begin{equation}
            T(t) = \frac{\sum\limits_q N_q(t)T_q(t)}{\sum \limits_q N_q(t)}, 
            \label{eq_temp_scal}
        \end{equation}
        so that it represents particle-weighted $T_\text{i}$.

        \begin{figure}[ht!]
            \includegraphics[width=0.52\linewidth]{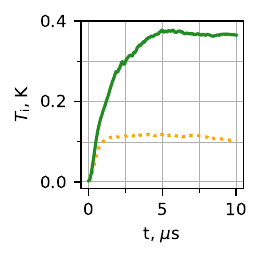}
            \caption{
                Time dependencies of $T_\text{i}$ for $T_\text{e0} = 10$~K (dashed) and $T_\text{e0} = 50$~K (solid).
                Here $N=2.5\cdot10^4$, $n_0~=~10^9$~cm$^{-3}$, $B = 100$~G.
            }
            \label{fig_temp}
        \end{figure}

            Fig. \ref{fig_temp} shows the dependence of the particle-weighted $T_\text{i}$ calculated using Eqs. (\ref{eq_temp_mesh}) and (\ref{eq_temp_scal}). 
        The data show that a higher initial electron temperature $T_\text{e0}$ results in a higher ion temperature $T_\text{i}$, which is contrary to the experimental results 
        reported in \cite{bib_pak2024preliminary}.

        Increasing the number of particles of the same charge, $N$, up to $10^5$ has no effect on this result.

            We also calculated the ion fluorescence spectra from the simulation data using the expressions given in \cite{CohenTannoudji1992}:
        \begin{equation}
            \begin{aligned}
                &\Delta_\text{L} = 2\pi f\\
                &\Delta_\text{i} = \Delta_\text{L} - kv_\text{i}\\
                &R_\text{i} = \frac{0.5\Gamma_0 s}{1 + s + \left(\frac{2\Delta_\text{i}}{\Gamma_0}\right)^2}, 
            \end{aligned}
        \end{equation}
        where $f$ is detuning frequency, $k$ is laser wavevector magnitude, $v$ is ion velocity component along the laser propagation direction, $s = 0.1$ is laser saturation parameter, 
        $\Gamma_0 = 11.49$~MHz \cite{bib_pak2024preliminary} is natural linewidth for Ca, and $R_\text{i}$ is photon scattering rate (fluorescence rate) of ion. Detuning range used is as wide as $6$~GHz.

        \begin{figure}[ht!]
            \includegraphics[width=\linewidth]{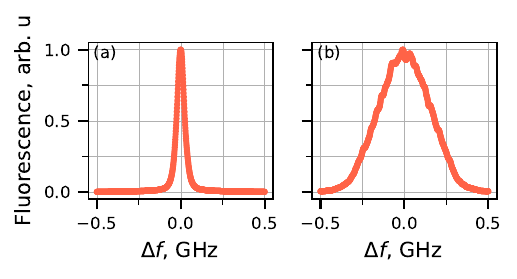}
            \caption{
                Ion spectra at $t = 10$~$\mu$s for:
                (a) across magnetic field lines; 
                (b) along magnetic field lines.
                Here $N=2.5\cdot10^4$, $n_0~=~10^9$~cm$^{-3}$, $B = 100$~G.
            }
            \label{fig_flour}
        \end{figure}
            Fig. \ref{fig_flour} shows the ion fluorescence spectra along the x and z directions. 
        Since the ion velocity distribution is anisotropic, the variance of the spectra depends on the direction. All ions were included in the spectrum calculation.

    \section{Discussion}

            The simulation of ultracold Ca plasma expansion in an external, constant, homogeneous magnetic field exhibits several notable features.

            Since the dependencies $\sigma^2(t)$ and $v_\text{exp}$ are in agreement with the experimental data, the model can be considered validated, at least with respect to the general plasma dynamics.
        In this regard, the simulation data on the ion and electron distributions complement the experimental data. 
        During plasma expansion in an external, constant, homogeneous magnetic field, an outer ion shell and an inner quasineutral plasma core are formed. 
        Moreover, despite the initial suppression of electron evaporation, the charge imbalance can reach significant values over time.

            However, there is a discrepancy between the simulation results and the experimental data on the ion temperature reported in \cite{bib_pak2024preliminary}. 
        It remains unclear whether the observed temperature is determined by the regularization parameter $r_0$, the number of particles, or both. 
        In this regard, it would be interesting to compare the experimental spectra obtained using laser-induced fluorescence with those obtained from molecular dynamics simulations.

           A separate issue concerns the appropriate criteria for comparing systems with large and small $N$. 
        Since systems with smaller $N$ are more prone to violations of electroneutrality, the $\Delta N$ inevitably affect their dynamics.

    \section{Acknowledgments}
            The research was supported by the Russian Science
        Foundation (Grant No. 23-72-10031-$\Pi$). The development
        of computational code work was supported by the Ministry of Science and Higher Education of the Russian Federation (State Assignment No. 075-00270-26-00).

    \nocite{*}
    \bibliographystyle{apsrev4-1}
    \bibliography{bibliography}

@article{bib_vikhrov2020simulated,
  title={Simulated expansion and ion front formation of ultracold plasma},
  author={Vikhrov, EV and Bronin, S Ya and Klayrfeld, AB and Zelener, BB and Zelener, BV},
  journal={Physics of Plasmas},
  volume={27},
  number={12},
  year={2020},
  publisher={AIP Publishing}
}

@article{bib_vikhrov2021ion,
  title={Ion wave formation during ultracold plasma expansion},
  author={Vikhrov, EV and Bronin, S Ya and Zelener, BB and Zelener, BV},
  journal={Physical Review E},
  volume={104},
  number={1},
  pages={015212},
  year={2021},
  publisher={APS}
}

@article{bib_bronin2023ultracold,
  title={Ultracold plasma expansion in quadrupole magnetic field},
  author={Bronin, S Ya and Vikhrov, EV and Zelener, BB and Zelener, BV},
  journal={Physical Review E},
  volume={108},
  number={4},
  pages={045209},
  year={2023},
  publisher={APS}
}

@article{vikhrov2025diocotron,
  title={Diocotron instability in ultracold plasma},
  author={Vikhrov, EV and Zelener, BB and Zelener, BV},
  journal={Physical Review E},
  volume={112},
  number={6},
  pages={065206},
  year={2025},
  publisher={APS}
}

@article{bib_zhang2008,
  title={Ultracold plasma expansion in a magnetic field},
  author={Zhang, X. and Fletcher, R. and Rolston, S. and Guzdar, P. and Swisdak, M.},
  journal={Physical review letters},
  volume={100},
  number={23},
  pages={235002},
  year={2008},
  publisher={APS}
}

@article{bib_sprenkle2022,
  title={Ultracold neutral plasma expansion in a strong uniform magnetic field},
  author={Sprenkle, T. and Bergeson, S. and Silvestri, L. and Murillo, M.},
  journal={Physical Review E},
  volume={105},
  number={4},
  pages={045201},
  year={2022},
  publisher={APS}
}

@article{bib_pak2024preliminary,
  title={Preliminary study of plasma modes and electron-ion collisions in partially magnetized strongly coupled plasmas},
  author={Pak, C. and Billings, V. and Schlitters, M. and Bergeson, S. and Murillo, M.},
  journal={Physical Review E},
  volume={109},
  number={1},
  pages={015201},
  year={2024},
  publisher={APS}
}

@article{lyubonko2012collective,
  title={Collective energy absorption of ultracold plasmas through electronic edge-modes},
  author={Lyubonko, Andrei and Pohl, Thomas and Rost, Jan-Michael},
  journal={New Journal of Physics},
  volume={14},
  number={5},
  pages={053039},
  year={2012},
  publisher={IOP Publishing}
}

@article{killian2007ultracold,
  title={Ultracold neutral plasmas},
  author={Killian, Thomas C and Pattard, T and Pohl, T and Rost, JM},
  journal={Physics reports},
  volume={449},
  number={4-5},
  pages={77--130},
  year={2007},
  publisher={Elsevier}
}

@book{CohenTannoudji1992,
  author    = {Cohen-Tannoudji, Claude and Dupont-Roc, Jacques and Grynberg, Gilbert},
  title     = {Atom-Photon Interactions: Basic Processes and Applications},
  publisher = {Wiley-Interscience},
  address   = {New York},
  year      = {1992},
  isbn      = {0-471-62556-6}
}

@article{Pilipp1971,
  author  = {Werner G. Pilipp},
  title   = {Expansion of an Ion Cloud in the Earth's Magnetic Field},
  journal = {Planetary and Space Science},
  volume  = {19},
  number  = {9},
  pages   = {1095--1119},
  year    = {1971},
  doi     = {10.1016/0032-0633(71)90107-3}
}

@article{warrens2023shockwaves,
  title={Wave steepening and shock formation in ultracold neutral plasmas},
  author={Warrens, MacKenzie and Inman, Nina and Gorman, Grant and Bennett, Husick and Bradshaw, Stephen and Killian, Tom},
  journal={Physics of Plasmas},
  volume={31},
  number={11},
  pages={113503},
  year={2024}
}
\end{document}